\documentstyle[prb,multicol,aps]{revtex}
\input{epsf}
\begin{document}
\draft

\title{On-chain electrodynamics of metallic (TMTSF)$_2 X$ salts: 
Observation of Tomonaga-Luttinger liquid response}

\author{A. Schwartz\cite{as_address}, M. Dressel\cite{md_address},
W. Henderson, and G. Gr\"uner}
\address{Department of Physics and Astronomy,
University of California, Los Angeles, California 90095-1547}
\author{V. Vescoli and L. Degiorgi}
\address{Laboratorium Festk\"{o}rperphysik, ETH-Z\"urich,
CH-8093 Z\"urich, Switzerland}
\author{T. Giamarchi}
\address{Laboratoire de Physique des Solides, Universit\'e Paris--Sud, 
B\^atiment 510, 91405 Orsay, France}
\date{Received 14 January 1998; To be published in Phys. Rev. B, 15 July 1998}
\maketitle

\begin{abstract}
We have measured the electrodynamic response in the metallic state of 
three highly anisotropic conductors, (TMTSF)$_2 X$, where $X$~=~PF$_6$, 
AsF$_6$, or ClO$_4$, and TMTSF is the organic molecule 
tetra\-methyl\-tetra\-seleno\-fulvalene. In all three cases we find 
dramatic deviations from a simple Drude response. The optical 
conductivity has two features: a narrow mode at zero frequency, with a 
small spectral weight, and a mode centered around 200~cm$^{-1}$, with 
nearly all of the spectral weight expected for the relevant number of 
carriers and single particle bandmass. We argue that these features are 
characteristic of a nearly one-dimensional half- or quarter-filled 
band with Coulomb correlations, and evaluate the finite energy mode in 
terms of a one-dimensional Mott insulator. At high frequencies ($\hbar\omega > 
t_\perp$, the transfer integral perpendicular to the chains), the 
frequency dependence of the optical conductivity $\sigma_1(\omega)$ is in 
agreement with calculations based on an interacting 
Tomonaga-Luttinger liquid, and is different from what is expected for an 
uncorrelated one-dimensional semiconductor. The zero frequency mode shows 
deviations from a simple Drude response, and can be adequately described 
with a frequency dependent mass and relaxation rate.
\end{abstract}

\pacs{PACS numbers: 78.20.-e, 71.10.Pm, 75.30.Fv}

\begin{multicols}{2}
\columnseprule 0pt
\narrowtext

\section{Introduction}
\label{sec:intro}

Since their first synthesis in the late 1970s, \cite{bechgaard_synthesis_first} 
the (TMTSF)$_2 X$ family of linear-chain organic conductors, 
and the closely related TMTTF family, has attracted continual 
attention. While the various broken symmetry ground states, including 
spin density waves, charge density waves, spin Peierls, and even 
superconductivity  have been extensively explored over the last two 
decades, \cite{gruner_revue_cdw,gruner_book_cdw,jerome_revue_1d} much of 
the current attention is now focused on the metallic state. These 
compounds have become one of the prototypical testing grounds for the 
study of the effects of electron--electron interactions in one-dimensional 
(1D) structures. In a strictly 1D interacting electron 
system, the Fermi liquid (FL) state is replaced by a state in which 
interactions play a crucial role, and which is generally referred to as a 
Tomonaga-Luttinger liquid (TLL). 

Fortunately, well defined techniques to treat such interactions exist in 
one dimension, and the physical properties of the TLL are 
well characterized and understood. 
\cite{emery_revue_1d,solyom_revue_1d,haldane_bosonisation} It is thus 
crucial to know how well the properties of the organic conductors are 
described by this TLL theory. The answer to this question is not clear from 
the outset because of the {\em quasi}-one-dimensional nature of the 
compounds. Due to interchain hopping, or interchain electron--electron or 
electron--phonon interactions, such materials are never strictly 
one-dimensional. In particular, although one expects the 1D 
theory to hold at high temperature, at lower temperatures the interchain 
hopping is expected to drive the system toward a more isotropic (2D or 
3D) behavior. These compounds are thus ideal candidates for a study of the 
dimensionality crossover between a non-Fermi liquid regime (TLL 
in this case) and a more conventional Fermi liquid state. Whether 
or not the low temperature phase is a FL, and the value of the 
expected crossover temperature, are issues which have been widely 
debated. Questions have also been raised about the importance of the 
interactions, due particularly to the important success of simple 
mean-field theories in explaining the properties of the low temperature 
condensed phases.

In the materials which we discuss in this paper, the
Bechgaard salts, charge transfer of one electron from every two TMTSF 
molecules leads to a quarter-filled (or
half-filled due to dimerization) hole band, thus enhancing the
importance of umklapp scattering. Such a state would be a Mott insulator 
in one dimension. As pointed out above, however, these materials are only 
quasi-one-dimensional, having two different finite interchain hopping 
integrals in the two transverse directions. Such an interchain coupling, 
however, becomes ineffective at high enough temperatures or frequencies, 
and thus we would expect the 1D physics to dominate in this regime.  
While in many of these materials the high temperature dc conductivity has 
an essentially metallic character, the finite frequency response is 
distinctly non-Drude.

In this paper we report on measurements of the electrodynamic response of 
three members of the (TMTSF)$_2 X$ family of charge-transfer salts 
($X$~=~PF$_6$, AsF$_6$, and ClO$_4$), with the goal of studying the nature 
of the metallic state in these quasi-1D conductors. In this regime, all 
three compounds exhibit behavior which cannot be described as the 
response of a simple Drude metal. In particular, two distinct features 
appear in the optical conductivity $\sigma_1(\omega)$, one at zero 
frequency and one at finite frequency. In Sec.~\ref{sec:expt} we discuss 
the sample growth techniques along with the experimental methods and 
results. Theoretical analysis of both modes which appear in the 
conductivity is presented in Sec.~\ref{sec:theory}. Finally in 
Sec.~\ref{sec:disclut} we conclude with a discussion of the experimental 
and theoretical results and the implications of our findings with regard 
to an understanding of the dynamics of quasi-1D systems.

A number of previous studies have investigated the electrodynamic 
response of various Bechgaard salts, but all covered only a portion of 
the spectral range presented in this paper, thus giving only a partial 
description of the complete response of these systems. 
\cite{Jacobsen81b,Kikuchi82,Jacobsen83a,Ng83,Challener84,Eldridge85b,Ng85a,Eldridge86a,Kornelsen87,Cao96} 
In addition, the small size of crystals 
previously available precluded the use of single crystals at all but the 
highest frequencies. As a result, all of the earlier work was done on 
composite samples, mosaics formed by aligning multiple crystals next to 
each other in order to produce an optical face. Indeed, our own initial 
work on these compounds was done in this way as 
well. \cite{donovan_pf6_reflectivity1,donovan_pf6_reflectivity2} However, as 
will be presented below, we have recently developed a method of growing 
single crystals of the Bechgaard salts with large transverse dimensions, 
yielding samples with appreciable optical faces in the {\em a-b} plane. 
Consequently, unlike any previous work, the results presented in this 
paper were all obtained on {\em single crystals}, even at the lowest 
frequencies. This has allowed us to avoid all of the complications and 
spurious results arising from the mosaic samples.

\section{Experimental methods and results}
\label{sec:expt}

\subsection{Sample preparation}
\label{sec:samp-prep}

Many of the measurements presented here were made possible by our ability
to grow large single crystals of the Bechgaard salts. The strong
anisotropy of the electronic structure tends to lead to long
needle-like crystals, with typical transverse dimensions of less than
0.5mm. In order to perform high quality optical measurements it is
necessary to have large crystal faces. This was historically obtained by
aligning multiple needle-like crystals into mosaics. This method
introduced uncertainty into the measurements due to misalignment of the
crystals, gaps between the crystals, and diffraction effects from the
composite sample.

The large single crystals used in this study were grown by the standard 
electrochemical growth technique,\cite{bechgaard_synthesis_first} but at 
reduced temperature ($0^{\circ}$C) and at low current densities. These 
conditions produce a slow growth rate and, over period of four to six 
months, single crystals up to $4\times 2.5 \times 1$~mm$^3$. These large, 
high-quality crystal faces in the {\em a--b$^{\prime}$} plane allowed us 
to perform reliable measurements of the electrodynamic response of these 
materials both parallel ($E\parallel a$) and perpendicular ($E\parallel 
b^{\prime}$) to the highly conducting chain axis down to low frequencies. 
The use of such crystals led to significant enhancement of the accuracy 
obtained for the optical reflectivity. In this paper we will focus on our 
results along the chains, some of which have been published 
previously.\cite{dressel_optical_tmtsf} Our findings regarding the 
conductivity perpendicular to the chains have been presented, 
\cite{degiorgi_tmtsf_perp} and will be discussed further in a future 
publication. \cite{Vescoli_SDWgap}

\subsection{Experimental techniques and analysis}
\label{sec:expt-tech}

By combining the results from various different spectrometers in the
microwave, millimeter, submillimeter, infrared, optical, and ultraviolet
frequency ranges, we have obtained the electrodynamic response of these
Bechgaard salts over an extremely broad range (0.1--10$^5$~cm$^{-1}$). 
In the optical range from 15~cm$^{-1}$ to $10^5$~cm$^{-1}$ standard
polarized reflectance measurements were performed employing four
spectrometers with overlapping frequency ranges. Two grating
spectrometers were employed at the highest frequencies: a McPherson
spectrometer in the ultraviolet, and a home-made spectrometer based on a
Zeiss monochromator in the visible. In the infrared spectral range two
Fourier transform interferometers were used with a gold mirror as the
reference. From the far-infrared up to the mid-infrared a fast scanning
Bruker IFS48PC spectrometer was employed, while in the far-infrared we
also made use of a Bruker IFS113v spectrometer with a mercury arc lamp
source and a helium-cooled germanium bolometric detector.

In the submillimeter spectral range (8--13~cm$^{-1}$), we have used a 
coherent source spectrometer \cite{volkov_coherent_spectrometer} based 
on backward wave oscillators, \cite{soohoo_book} high power, tunable, 
monochromatic light sources with a broad bandwidth. These oscillators 
operate at frequencies below those accessed by a typical infrared 
spectrometer, and above the millimeter wave range, thus filling in an 
admittedly narrow, but important, gap in energy between these two more 
widely studied regions. This spectrometer was originally designed for 
transmission measurements and we have reconfigured it for reflection 
measurements on highly conducting bulk samples. The reflection 
coefficient is obtained by comparing the signals reflected from the 
sample and from a polished aluminum reference mirror, where the 
reflectivity of aluminum was calculated from the dc conductivity 
$\sigma_{\rm dc}$ by the Hagen-Rubens relation\cite{ziman_solid_book}
\begin{equation}
R(\omega)=1-\left(\frac{2\omega}{\pi\sigma_{\rm dc}}\right)^{1/2}.
\label{eq:hr}
\end{equation}
This expression is valid for $\sigma_1 \gg
|\sigma_2|$, and in this limit $\sigma_1(\omega)\simeq\sigma_{\rm dc}$.

At all frequencies up to and including the mid-infrared, we placed the samples
in an optical cryostat and measured the reflectivity as a function of
temperature between 5K and 300K. In the case of (TMTSF)$_2$PF$_6$ these
reflectivity data were combined with previous measurements in the
microwave and millimeter wave spectral range
\cite{donovan_pf6_reflectivity1,donovan_pf6_reflectivity2}
which were made by the use of a resonant cavity perturbation
technique.\cite{klein_cavity} In this method, the surface impedance
$\hat{Z}_S=R_S+iX_S$ of the material can be measured by placing a needle
shaped crystal in an anti-node of either the electric or magnetic field
of a cylindrical cavity in the TE$_{011}$ mode and measuring the change
in width ($\Delta\Gamma$) and center frequency ($\Delta f$) of the
resonance. It is then possible to calculate both the surface resistance
$R_S$ and the surface reactance $X_S$ as follows:
\begin{equation}
R_S=Z_0\frac{\Delta\Gamma}{2f_0\zeta} \hspace*{0.5cm} {\rm and}
\hspace*{.5cm} X_S=Z_0\frac{\Delta f}{f_0\zeta},
\end{equation}
where $Z_0={4\pi}/{c}=4.19\times 10^{-10}$~s/cm is the impedance of free
space ($Z_0=377~\Omega$ in SI units). The resonator constant $\zeta$ can
be calculated from the geometry of the cavity and the sample.
\cite{klein_cavity}

The complex conductivity ($\hat{\sigma}=\sigma_1+i\sigma_2$) can be calculated
from the surface impedance using $\hat{Z}_S=Z_0\sqrt{\omega/4\pi i\hat{\sigma}}$, 
and the absorptivity $A$ is given by the relation
\begin{equation}
A =1-R = \frac{4R_S}{Z_0}\left(1+\frac{2R_S}{Z_0}+
\frac{R_S^2+X_S^2}{Z_0^2}\right)^{-1},
\label{eq:mw-abs}
\end{equation}
where $R$ is the reflectivity. In the limit  $R_S, |X_S| \ll
Z_0$, Eq.~(\ref{eq:mw-abs}) reduces to
\begin{equation}
A\approx \frac{4R_S}{Z_0}.
\label{eq:mw-abs-hr}
\end{equation}
By using Eq.~(\ref{eq:mw-abs-hr}) it is possible to combine these
microwave and millimeter wave cavity data with all of the higher
frequency reflectivity data. In general, the reflectivity is a complex
function $\hat{r}(\omega)=|r(\omega)|e^{i\phi(\omega)}$. The reflectivity
$R$ which we measure is actually the square of this quantity,
$R=|\hat{r}|^2$. In order to obtain the phase $\phi$ we have performed a
Kramers-Kronig analysis on the reflectivity spectra: \cite{wooten_book}
\begin{equation}
\phi(\omega) =
\frac{\omega}{\pi}\int_{0}^{\infty}\frac{\ln{[R(x)]} -
\ln{[R(\omega)]}}{\omega^2 - x^2}dx,
\label{KK}
\end{equation}
where the $\ln{[R(\omega)]}$ term has been added to the standard form in
order to remove the singularity at $x=\omega$. It has no effect on the
integral because $\int_0^\infty (\omega^2 - x^2)^{-1}dx=0$. Because this
integral extends from zero to infinity, it is necessary to make suitable
high and low frequency extrapolations to the measured reflectivity data.
We have chosen to use a power law at high frequencies ($R(\omega)\propto
1/\omega^4$) and a Hagen-Rubens extrapolation, as given in
Eq.~(\ref{eq:hr}), to zero frequency. From $R(\omega)$ and $\phi(\omega)$
it is then possible to calculate the components of the complex optical
conductivity $\hat{\sigma}(\omega)$.

\begin{figure}[htb]
\begin{center}
\leavevmode
\epsfxsize=7cm
\epsfbox[75 65 460 775]{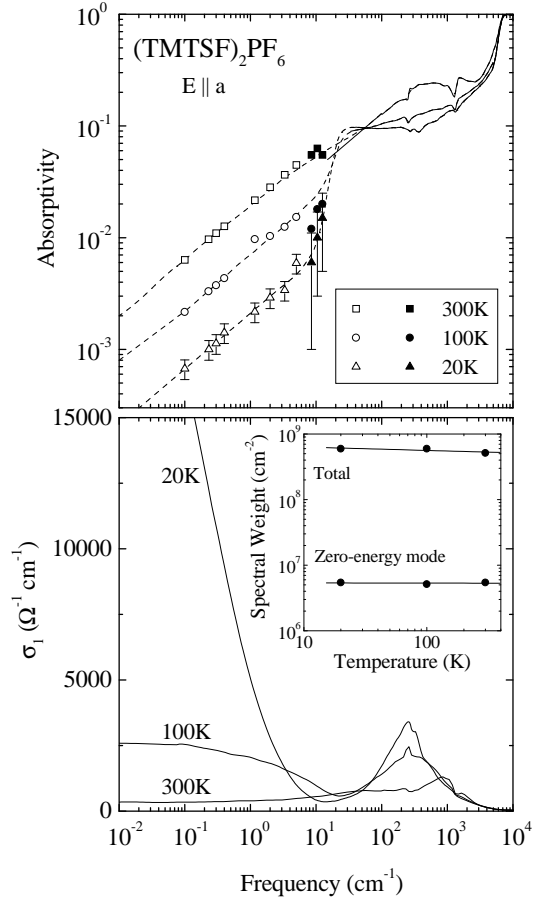}
\end{center}
\caption{\label{fig:pf6-as}
The measured absorptivity (a) and conductivity (b) of (TMTSF)$_2$PF$_6$
at 300K, 100K, and 20K, for $E\parallel a$. The open symbols were obtained
by the cavity perturbation technique. The solid symbols are
from submillimeter reflectivity measurements. The solid lines are
infrared through ultraviolet reflectivity data. The dashed lines show the
spectra which were used as the input for the KK calculations. The inset
in (b) shows the total integrated spectral weight and the fraction in
the zero-energy mode, both as functions of temperature, demonstrating
that the spectral weight is not redistributed, and that only about 1\%
is in the Drude-like mode.}
\end{figure}

\subsection{Experimental results}
\label{sec:expt-results}

The dc resistivities of the Bechgaard salts (TMTSF)$_2 X$ ($X$=PF$_6$, AsF$_6$, 
ClO$_4$) measured along their highly conducting chain axes exhibit metallic 
temperature dependence at high temperatures.
\cite{jerome_revue_1d,jerome_organic_review,samboni_clo4_dc} 
In addition, all three compounds exhibit a sharp phase transition near 10~K, 
and a thermally activated behavior of the resistivity below this 
temperature. It has been established that the ground state in each of 
these materials is a spin density wave (SDW), \cite{clo4_dc_note}  
with transition temperatures of 12K for the PF$_6$ and AsF$_6$
compounds, and 6K for ClO$_4$. \cite{gruner_revue_cdw,gruner_book_cdw} 
The insulating behavior results from an opening of a gap in the single 
particle excitation spectrum at the Fermi level due to electron--electron 
interactions.

\begin{figure}[htb]
\begin{center}
\leavevmode
\epsfxsize=7cm
\epsfbox[75 65 460 775]{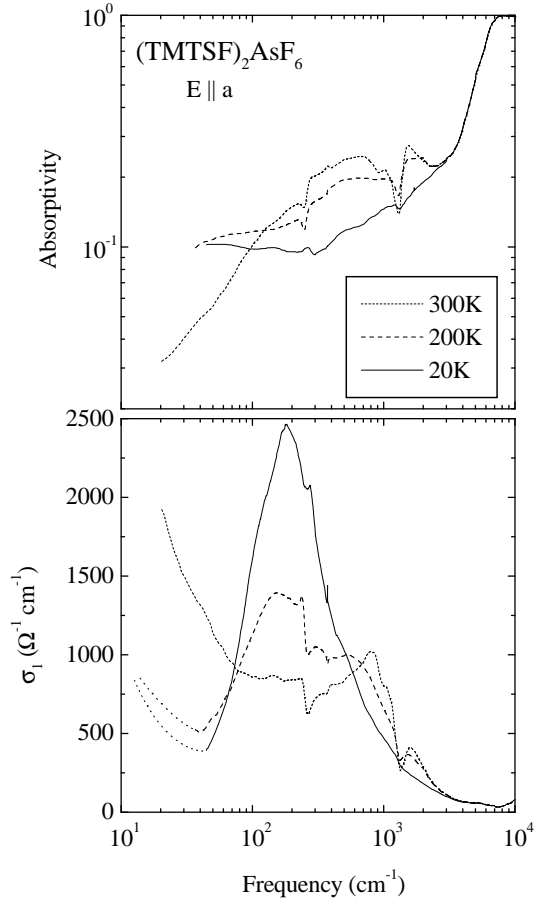}
\end{center}
\caption{\label{fig:asf6-as}
The measured absorptivity (a) and conductivity (b) of (TMTSF)$_2$AsF$_6$
at 300K, 200K, and 20K, for $E\parallel a$. In the absence of any low
frequency data, we are unable to resolve the low frequency mode. The
dotted lines at the low frequency end of the 20K and 200K spectra in (b)
are drawn only to indicate the necessity of such a mode in order to meet
the dc conductivity values.}
\end{figure}

\begin{figure}[htb]
\begin{center}
\leavevmode
\epsfxsize=7cm
\epsfbox[75 65 460 775]{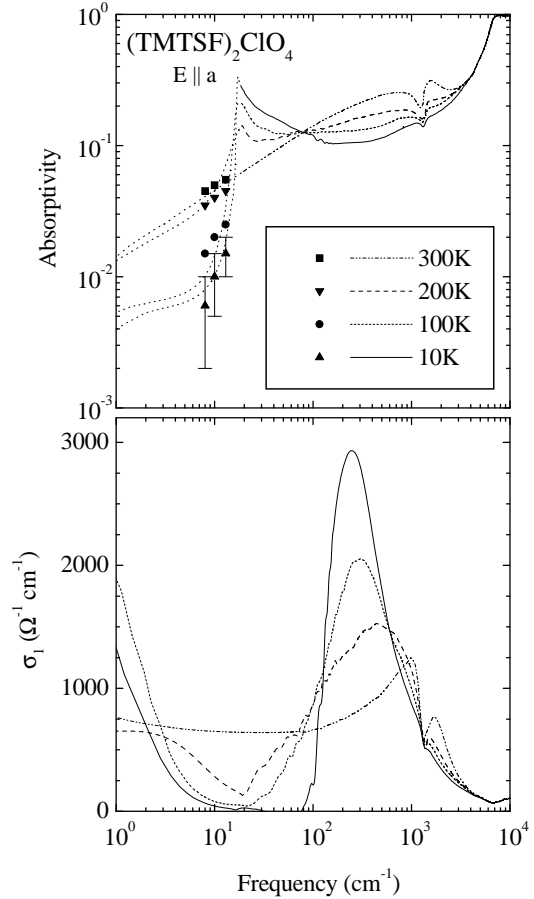}
\end{center}
\caption{\label{fig:clo4-as}
The measured absorptivity (a) and conductivity (b) of (TMTSF)$_2$ClO$_4$
at 300K, 200K, 100K, and 10K, for $E\parallel a$. The solid symbols are data
from submillimeter reflectivity measurements. Below these frequencies,
Hagen-Rubens extrapolations were made to meet the dc conductivity values.}
\end{figure}

Figures~\ref{fig:pf6-as}--\ref{fig:clo4-as} show the measured
absorptivities as a function of frequency at various 
temperatures for each of the compounds, and
the conductivities calculated by use of the Kramers-Kronig relation, as
described above. Unlike the response of a simple metal, the frequency
dependent conductivities of these compounds cannot be described by a
single Drude term of the form
\begin{equation}
\hat{\sigma}(\omega)=\frac{\sigma_0}{1-i\omega\tau}=\frac{\omega_p^2}{4\pi}
\frac{1}{\Gamma-i\omega},
\label{eq:sim-drude}
\end{equation}
where $\sigma_0=ne^2\tau/m_b$ is the dc conductivity, $\omega_p^2=4\pi
n e^2/m_b$ is the plasma frequency, and $\Gamma=1/\tau$ is the scattering
rate of the carriers with bandmass $m_b$ and number density $n$.
Instead the spectrum exhibits two distinct features. The
high dc conductivities are associated with a very narrow
mode, centered at zero energy (ZE). In addition there is a second 
finite energy (FE) excitation centered near 200~cm$^{-1}$. 
The combined spectral weight of these two modes is given by
\begin{equation}
\int\sigma_1^{\rm ZE}(\omega)\,d\omega + \int \sigma_1^{\rm
FE}(\omega)\,d\omega = \frac{\pi
ne^2}{2m_b}=\frac{\omega^2_p}{8},
\label{eq:spec-wt}
\end{equation}
and leads to a total plasma frequency $\omega_p/(2\pi c)=1.1\times 
10^4~{\rm cm}^{-1}$. As is shown in the inset of Fig.~\ref{fig:pf6-as}b, 
this total spectral weight is independent of temperature, as is expected. 
In addition, the value is in full agreement with that obtained from the 
known carrier concentration $n=1.24\times 10^{21}~{\rm cm}^{-3}$ and a 
bandmass $m_b\approx m_e$. \cite{jerome_revue_1d}

Figure~\ref{fig:pf6-as}b also demonstrates that most of the spectral
weight resides in the finite energy mode, with only about 1\% in
the zero energy mode, independent of temperature. This value can be
calculated in two ways. The contribution of the finite energy mode can be
subtracted from the conductivity, and the remaining zero energy mode can
be integrated. This approach gives $\omega_p^{\rm ZE}/(2\pi c)=
1000\pm100~{\rm cm}^{-1}$. Alternately, the zero crossing of the
dielectric constant, when corrected for the higher frequency
contributions from the finite energy mode, gives a plasma frequency of
approximately $1\times 10^3{\rm cm}^{-1}$, at all temperatures where this
mode is clearly defined. \cite{dressel_optical_tmtsf} This rather unusual
two-structure response is the focus of the work presented here. We
discuss the structure and possible origins of these two excitations
separately in the following sections, although it is our belief that they
result from the same underlying interactions within these materials.

\section{Theoretical analysis} \label{sec:theory}

\subsection{Overview} \label{sec:over}

Before turning to the various excitations observed in the optical 
spectrum, we discuss the overall features: the appearance of both a 
zero-energy and finite-energy excitation, with most of the spectral 
weight associated with the latter. This unusual behavior of the 
conductivity makes it unlikely that it can be described by a conventional 
explanation in terms of a weakly interacting (or Fermi liquid-like) 
description.\cite{gorkov_sdw_tmtsf} Such a description would hardly be 
consistent with the fact that $99\%$ of the spectral weight is contained 
in the finite frequency mode around 200~cm$^{-1}$.

In order to analyze this structure one has to remember that the 
(TMTSF)$_2 X$ family is a good realization of a quasi-one-dimensional 
electron system with some hopping between chains, and to compare the 
behavior of two isostructural groups of organic conductors, the 
(TMTTF)$_2 X$ and (TMTSF)$_2 X$ salts. The (TMTSF)$_2 X$ salts have a larger 
interchain hopping than the (TMTTF)$_2 X$ salts due to the larger overlap 
of the molecular orbitals in the direction perpendicular to the chains. 
Various estimates of the transfer integrals in the TMTSF family lead to 
the following values along the three directions: \cite{jerome_revue_1d} 
$t_a$, $t_b$, $t_c$ = 250meV, 25meV, 1meV. Although the chains are 
coupled by an interchain hopping $t_\perp$, such an interchain coupling 
becomes ineffective at high temperatures or frequencies, and above 
a certain energy scale $\text{Min}(\omega,T) > E_{\text{cr}}$ the system 
possesses a one-dimensional character. The naive value for 
$E_{\text{cr}}$  is $E_{\text{cr}} \sim t_\perp$, but interactions can 
renormalize this value downwards, leading to a wider one-dimensional 
regime.\cite{bourbonnais_couplage,boies_hopping_general} 

In both cases, due to full charge transfer from the organic molecule to 
the counter ions, the TMTTF or TMTSF stacks have a  quarter-filled hole 
band. There is also a moderate dimerization, which is somewhat more 
significant for the TMTTF family. Therefore, depending on the importance 
of this dimerization, the band can be described as either half-filled 
(for a strong dimerization effect) or quarter-filled (for weak 
dimerization). Due to the commensurate filling, a strictly 
one-dimensional model is expected to lead to a Mott insulating behavior.  
Indeed, the (TMTTF)$_2 X$ salts, with $X$~=~PF$_6$ or Br, are insulators at 
low temperatures \cite{jerome_revue_1d} with a substantial single 
particle gap ($\Delta\approx 2000$~cm$^{-1}$ for the PF$_6$ salt). For 
these compounds the gap is so high that the interchain hopping is 
suppressed by the insulating nature of the 1D phase and is not relevant.

For the TMTSF family, however, one expects a stronger competition between 
the Mott gap and the interchain hopping. In a very crude way, the 
interchain hopping could be viewed as an effective doping leading to 
deviations from the commensurate filling (which is insulating). The 
optical features we observe are close to those which have been calculated 
for a {\em doped} one dimensional Mott insulator: 
\cite{giamarchi_umklapp_1d,giamarchi_mott_shortrev,mori_mott_1d} a Mott 
gap and a zero-energy mode for small doping levels. Of course for the low 
energy mode, this is an oversimplified view, since the interchain 
hopping  makes the system two dimensional, and the low energy feature is 
unlikely to be described by a simple one dimensional theory.

We now discuss in detail the two distinct features observed in the 
optical conductivity. While it is clear from 
Figs.~\ref{fig:pf6-as}--\ref{fig:clo4-as} that these two features develop 
progressively as the temperature is lowered, we will focus our attention 
on the lowest temperatures (10--20K), just above the phase transitions to 
broken-symmetry ground states.

\subsection{Finite-energy mode}
\label{sec:finite-energy}

As discussed above, the naive upper limit for the crossover from 1D to 2D 
or 3D nature is $E_{\rm cr}\sim t_\perp \approx 200-300$cm$^{-1}$, and 
therefore the peak structure in the conductivity should be well described 
by a purely one-dimensional theory. In such a 1D regime the 
effects of electron--electron interactions are particularly important and 
lead to the formation of a non-Fermi liquid state, the so-called 
Tomonaga-Luttinger liquid (TLL). 
\cite{emery_revue_1d,solyom_revue_1d,haldane_bosonisation} Such a state 
is characterized by an absence of single electron-like quasiparticles and 
by a non-universal decay of the various correlation functions. All of the 
excitations of the system, even the single particle excitations, can be 
described in terms of charge density fluctuations, whose energy is given 
by
\begin{equation} \label{quadra}
H_0 = \frac1{2\pi} \int dx \; u_\rho K_\rho (\pi\Pi_\rho)^2 +
                 \frac{u_\rho}{K_\rho} (\nabla \phi)^2,
\end{equation}
where $\nabla \phi = \rho(x)$, the charge density, and $\Pi$
is the conjugate momentum to $\phi$. All of the interaction effects are
hidden in the parameters $u_\rho$ (the velocity of charge excitations)
and $K_\rho$ (the Luttinger liquid exponent controlling the decay of all
correlation functions). Initially derived for interactions much weaker
than the bandwidth, \cite{emery_revue_1d,solyom_revue_1d} this description
has been proven to be valid for an arbitrary one-dimensional interacting
system, \cite{haldane_bosonisation} provided one uses the proper $u_\rho$
and $K_\rho$. The TLL description can be viewed as an effective
low energy description of the one-dimensional electron gas, which is
reminiscent of FL theory as a low energy description of the
three-dimensional interacting electron gas (with quite different
physics). The above Luttinger liquid parameters can be viewed as the
1D equivalents of the Landau parameters of a FL. In a general way, 
$K_\rho = 1$ is the noninteracting point, with $K_\rho > 1$ 
corresponding to attraction and $K_\rho < 1$ to repulsion.

Depending on the interaction strength, the TLL may be unstable for 
commensurate fillings, causing the system to become a Mott insulator, 
with a charge gap. \cite{emery_revue_1d,solyom_revue_1d,lieb_hubbard_exact} 
The mechanism leading to this insulating state is now well understood and 
can be easily described within the TLL formalism. When the filling 
is commensurate, interactions give rise to additional contributions to the 
TLL Hamiltonian. These are the so-called umklapp processes of the form
\begin{equation} \label{um1n}
H_{1/2n} = g_{1/2n} \int dx \cos \left(n \sqrt8 \phi_\rho(x)\right),
\end{equation}
where $n$ is the order of the commensurability
($n=1$ for half filling -- one particle per site 
\cite{emery_revue_1d,solyom_revue_1d,giamarchi_umklapp_1d};
$n=2$ for quarter filling -- one particle every
two sites, and so on 
\cite{giamarchi_mott_shortrev,giamarchi_curvature,schulz_mott_revue}).
The coupling constant $g_{1/2n}$ is the umklapp
process corresponding to the commensurability $n$. The commensurability
of order $n$ is relevant provided $n^2 K_\rho < 1$ (the higher the
commensurability the more repulsive the interaction has to be). When the
umklapp term is relevant it opens a gap in the charge spectrum of the
form
\begin{equation} \label{legapch}
\Delta_\rho \sim W \left(\frac{g_{1/2n}}{W} \right)^{1/(2-2n^2K_\rho)},
\end{equation}
for $g_{1/2n}$ much less than the bandwidth $W$.

This is the relevant situation to describe the organic compounds because 
the band is quarter-filled. A quarter-filling umklapp thus exists with a 
coefficient of order $g_{1/4}\sim W (U/W)^3$, for a typical interaction 
of order $U$ and a bandwidth of order $W$. However, because the chains 
are slightly dimerized, a half-filled umklapp is also generated with a 
value on the order of $g_{1/2}\sim U (D/E_F)$  where $D$ is the 
dimerization gap.\cite{emery_umklapp_dimerization} Which umklapp term 
dominates depends, of course, on the strength of the interactions, on 
$K_\rho$ (also controlled by the interactions), and on the magnitude of 
the dimerization gap. This 1D insulator picture thus 
provides a good description of the gross features observed in the 
conductivity. The peak corresponds to the Mott gap and would also be 
consistent with finding most (or {\em all}, if the system was a purely 
one-dimensional Mott insulator) of the spectral weight in the finite 
frequency peak.

\begin{figure}[htb]
\begin{center}
\leavevmode
\epsfxsize=7cm
\epsfbox[75 240 555 760]{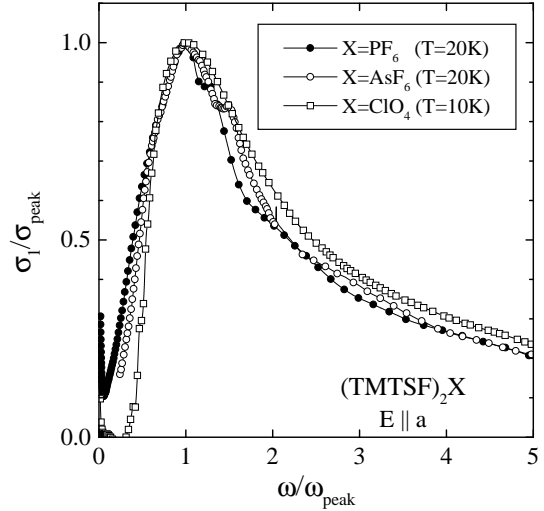}
\end{center}
\caption{\label{fig:rescaled}
The frequency dependent conductivities of (TMTSF)$_2 X$ (X=PF$_6$, AsF$_6$,
ClO$_4$). Both axes have been normalized to the peak of the finite energy
mode near 200~cm$^{-1}$. The universal behavior of $\sigma_1(\omega)$ is 
evident.}
\end{figure}

\begin{figure}[htb]
\begin{center}
\leavevmode
\epsfxsize=7cm
\epsfbox[60 240 555 760]{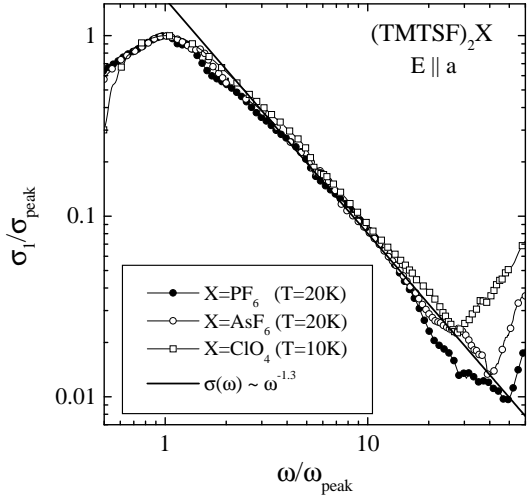}
\end{center}
\caption{\label{fig:powerlaw}
The normalized conductivities from Fig.~\protect\ref{fig:rescaled}
shown on a log--log scale to demonstrate
the power law frequency dependence of the conductivities
above the finite energy peak. The solid line
shows a fit of the form $\sigma(\omega)\sim\omega^{-\nu}$ as discussed in
the text. We find that for all three compounds, $\nu=1.3 \pm 0.1$. }
\end{figure}

Because a complete description of the dynamic conductivity in such a 
one-dimensional Mott insulator exists, 
\cite{giamarchi_umklapp_1d,giamarchi_mott_shortrev,giamarchi_curvature} 
one can make {\em quantitative}\ comparisons of the data to the theoretical 
predictions. The salient feature is a nonuniversal power law behavior 
controlled by the interactions for $\omega \gg \Delta_\rho$:
\begin{equation} \label{powerlaw}
\sigma(\omega) \sim \omega^{4 n^2 K_\rho - 5}.
\end{equation}
Besides providing a good test of the one-dimensional theory, a fit of the 
data to the form of Eq.~(\ref{powerlaw}) gives direct access to the 
Luttinger liquid exponent $K_\rho$. We will come back to the 
determination of $K_\rho$ and the physical consequences in 
Section~\ref{sec:disclut}, and confine the current discussion to a direct 
analysis of the data. Indeed when rescaled by the Mott gap, the three 
systems ($X$=ClO$_4$, PF$_6$, and AsF$_6$) exhibit remarkably similar 
behavior, as demonstrated in Fig.~\ref{fig:rescaled}. A log--log plot of 
the data for the three different systems is shown in 
Fig.~\ref{fig:powerlaw}. The agreement with the prediction of the 
one-dimensional theory is quite good over more than one decade in 
frequency, leading to an exponent $4 n^2 K_\rho - 5 \approx -1.3$. The 
observed power law in the frequency dependence is also in very reasonable 
agreement with the observed temperature dependence of the resistivity, which is 
found to be roughly linear above 100K once thermal expansion has been 
taken into account. \cite{jerome_organic_review} Indeed in a purely 
1D model, the temperature dependence of the resistivity is given by 
\cite{giamarchi_umklapp_1d,giamarchi_mott_shortrev,giamarchi_curvature}
\begin{equation}
\rho(T) \sim T^{4 n^2 K_\rho - 3}
\end{equation}
and thus the frequency and temperature dependence should be related by
\begin{equation} \label{eq:consist}
\sigma(\omega) \sim \omega^{-\nu} \qquad \rho(T) \sim T^{2-\nu}
\end{equation}
with no undetermined adjustable parameter. Although a precise determination
of the temperature exponent is difficult because of thermal
expansion and the possibility of other sources of scattering,
Eq.~(\ref{eq:consist}) is  again consistent with the data.

It thus seems very natural to interpret the high frequency part of the 
optical conductivity in terms of a Mott insulator. The observed Mott gap 
would be rather large ($\sim 100 \text{cm}^{-1}$ because the optical gap 
is twice the thermodynamic one). Due to the apparent contradiction of 
having a rather large Mott gap and a good metallic dc conductivity, it 
was proposed that the  peak structure was due to the dimerization gap $D$ 
itself. \cite{pedron_tmtsf_optics,mila_tmtsf_optics}. This would indeed 
be the case for an extremely strong (nearly infinite) repulsion, with the 
quarter-filled band being transformed into a half-filled band of (nearly 
non-interacting) spinless fermions. It was then argued that the real 
charge gap of the problem was smaller, on the order of $50$K. However, 
such an interpretation fails to reproduce the observed frequency 
dependence above the peak in conductivity. Using the same type of 
analysis for the conductivity as in Ref.~\onlinecite{giamarchi_umklapp_1d}, 
and attributing the peak to the dimerization gap, results in 
$\sigma(\omega)\sim 1/\omega^3$ for $\omega \gg \Delta$, 
as in a simple semiconductor (corresponding to 
nearly free spinless fermions). The observed power-law (see 
Fig.~\ref{fig:powerlaw}) differs significantly from this prediction, 
making such an interpretation of the data very unlikely.

As mentioned in Section~\ref{sec:over}, there is in fact no contradiction 
between a good metallic dc conductivity and a large Mott gap, provided 
that the system is doped slightly away from commensurate filling. 
\cite{giamarchi_mott_shortrev} Indeed, in a very crude way this is what 
seems to be observed here, with the ``Drude'' peak containing only one 
hundredth of the carriers. Although no real doping exists from a 
chemistry point of view, one could attribute such a deviation from 
commensurability to the effects of interchain hopping. If single particle 
hopping between chains is relevant, small deviations from commensurate 
filling due to the warping of the Fermi surface exist, and should lead to 
effects equivalent to real doping on a single chain. Of course, such a 
picture is only a poor man's way of viewing the low frequency structure. 
Since the interchain hopping is relevant, the low frequency peak 
should in principle be described by a full two-dimensional theory (of 
interacting fermions). Due to the complexity of such a problem, it is 
thus interesting to compare, from a purely phenomenological point of 
view, the shape of the observed Drude peak with FL predictions 
as well as with the predictions of the naive one-dimensional theory of a 
doped Mott insulator. Such a comparison is performed in the next section.

\subsection{Zero-energy mode}
\label{sec:zero-energy}

\begin{figure}[htb]
\begin{center}
\leavevmode
\epsfxsize=7cm
\epsfbox[90 235 555 760]{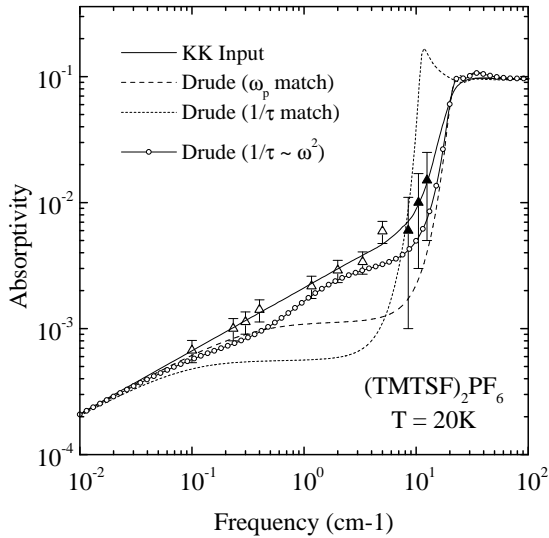}
\end{center}
\caption{\label{fig:abs-low}
The low frequency absorptivity of (TMTSF)$_2$PF$_6$ at $T=20$K. The triangles
are the same data shown in Fig.~\protect\ref{fig:pf6-as}(a) and the solid
line is the interpolation used as input into the Kramers-Kronig
calculation to arrive at the conductivity in
Fig.~\protect\ref{fig:pf6-as}(b). The dashed lines show two attempts to fit
this part of the spectrum with a simple
Drude of the form given in Eq.~({\protect\ref{eq:sim-drude}}), and the
line with open circles is a fit with a frequency dependent scattering
rate and effective mass, as described in the text. }
\end{figure}

\begin{figure}[htb]
\begin{center}
\leavevmode
\epsfxsize=7cm
\epsfbox[45 235 555 760]{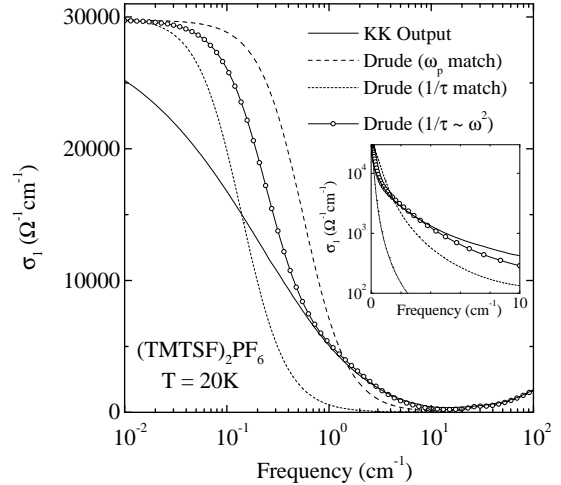}
\end{center}
\caption{\label{fig:sig-low}
The low frequency conductivity of (TMTSF)$_2$PF$_6$ at $T=20$K. The solid line
is the output of the Kramers-Kronig calculation. The two dashed lines
show the two simple Drude fits, corresponding to the dashed lines in
Fig.~{\protect\ref{fig:abs-low}}. The line with the open circles is a fit
using a generalized Drude form with a frequency dependent scattering rate
and mass. The inset shows the same spectra on a linear frequency scale. }
\end{figure}

The model presented in Sec.~\ref{sec:theory} predicts the appearance of a 
delta function dc conductivity in a 1D Hubbard system which is doped away 
from half-filling. As we have demonstrated, the finite frequency mode 
which develops in these materials is in agreement with such a model, and 
we do indeed see a corresponding peak in the conductivity at zero 
frequency. This extremely narrow resonance leads to the high dc 
conductivities seen in these materials, yet it accounts for only 1\% of 
the total spectral weight, as demonstrated in the inset of 
Fig.~\ref{fig:pf6-as}b. A similar peak, with an even smaller fraction of 
the spectral weight, has recently been reported in (TMTSF)$_2$ClO$_4$. 
\cite{Cao96} Qualitatively, this mode looks like a narrow Drude 
response, of the form given in Eq.~(\ref{eq:sim-drude}), but our data 
on the (TMTSF)$_2$PF$_6$ compound indicate that this is not 
the case. The two dashed lines in Figs.~\ref{fig:abs-low} and 
\ref{fig:sig-low} show fits to the low frequency data with a simple Drude 
form. In these two fits, either the spectral weight ($\nu_p = 
1000$~cm$^{-1}$) or width (1/$\tau=0.14$~cm$^{-1}$) of the mode was 
matched and the other was chosen in order to make the fit meet the 
measured dc conductivity value of $3\times 10^4$~($\Omega$cm)$^{-1}$. It 
is clear that neither gives a satisfactory fit to either the conductivity 
or absorptivity.

Although this analysis depends strongly on the microwave cavity 
perturbation data between 0.1 and 5~cm$^{-1}$, this technique is well 
developed and produces reliable data even for such highly conducting 
materials. \cite{klein_cavity} In addition, 
at 9~GHz a number of other measurements 
\cite{walsh_tmtsf_optics,janossy_tmtsf_optics,buravov_tmtsf_optics,javadi_tmtsf_optics} 
of the surface resistance have been made and all 
agree with our results within the experimental uncertainty shown on 
Fig.~\ref{fig:abs-low}. Musfeldt {\em et al.} 
\cite{musfeldt_tmtsf_optics} have also measured this compound in a 
16.5~GHz cavity. From their data we have calculated an absorptivity of 
approximately $4\times 10^{-3}$, which is significantly higher than we 
have found, in even greater disagreement with the two simple Drude fits. 
Finally, at 60~GHz we were able to measure both $R_S$ and $X_S$ in the 
maxima of both the electric and magnetic fields 
\cite{klein_cavity} and have found that 
within experimental error they are equal to each other and to the dc 
resistivity at 20K. \cite{donovan_pf6_reflectivity2} We are certain that 
the discrepancies between the measured data and the simple Drude fits, 
especially at the higher frequencies, are outside of the experimental 
uncertainty.

Similarly, it is clear that neither the low energy mode nor the conductivity 
below the Mott gap can be described {\em quantitatively}\ by a simple 
one-dimensional theory, even taking into account a phenomenological 
doping. Indeed, in a purely 1D theory the conductivity 
$\sigma(\omega)$ should grow as $\sigma(\omega)\sim \omega^3$ between the 
Drude peak and the Mott peak, 
\cite{giamarchi_mott_shortrev,giamarchi_curvature} but such a power is 
not observed experimentally. In addition, in a doped one-dimensional 
system the width of the $\omega=0$ peak remains extremely narrow since 
all electron--electron scattering that can lead to dissipation has been 
shifted to higher energy.

In order to obtain a satisfactory description of these low frequency data,
it thus seems necessary to turn to a Fermi liquid (FL) picture. 
\cite{pines_nozieres_book}
Let us introduce a complex frequency dependent scattering
rate $\hat{\Gamma}(\omega)=\Gamma_1(\omega)+i\Gamma_2(\omega)$ into the
standard Drude form of Eq.~(\ref{eq:sim-drude}). If we define the 
dimensionless quantity
$\lambda(\omega)=-\Gamma_2(\omega)/\omega$ then the complex 
conductivity can be written as
\begin{equation}
\hat{\sigma}(\omega)=\frac{\omega_p^2}{4\pi}
\frac{1}{\Gamma_1(\omega)-i\omega(m^*(\omega)/m_b)}
\label{eq:gen-drude2}
\end{equation}
where $m^*/m_b=1+\lambda(\omega)$ is the frequency dependent enhanced mass.
By rearranging Eq.~(\ref{eq:gen-drude2}) we can write expressions for
$\Gamma_1(\omega)$ and $m^*(\omega)$ in terms of $\sigma_1(\omega)$ and
$\sigma_2(\omega)$ as follows:
\begin{equation}
\Gamma_1(\omega)=\frac{\omega_p^2}{4\pi}
\frac{\sigma_1(\omega)}{|\hat{\sigma}(\omega)|^2}
\label{eq:gam-w}
\end{equation}
\begin{equation}
\frac{m^*(\omega)}{m_b}=\frac{\omega_p^2}{4\pi}
\frac{\sigma_2(\omega)/\omega}{|\hat{\sigma}(\omega)|^2}.
\label{eq:mstar-w}
\end{equation}
Due to causality, \cite{allen_drude} $\Gamma_1(\omega)$ and $m^*(\omega)$ 
are related through the Kramers-Kronig relation. \cite{wooten_book} Such 
analysis, which  allows us to look for 
interactions which would lead to frequency dependent scattering rates, 
has been used before in studying the response of heavy fermion 
compounds \cite{sulewski_drude_fit} and high temperature 
superconductors. \cite{ruvalds_nested_fl}

The Landau FL theory \cite{pines_nozieres_book} predicts that
the scattering rate due to electron--electron interactions in three 
dimensions should be quadratic in both temperature and
frequency. \cite{ruvalds_nested_fl,gurzhi_scattering_fl}
In order to examine the shape of the Drude-like peak observed in the 
data, we have adopted the following phenomenological forms of $\Gamma(\omega)$ and 
$m^*(\omega)$, used by Sulewski~{\em et al.} in their study of the FL 
behavior of the heavy fermion compound UPt$_3$: \cite{sulewski_drude_fit}
\begin{equation}
\Gamma_1(\omega)= \Gamma_0 + \frac{\lambda_0 \alpha \omega^2} 
{1+ \alpha^2\omega^2}
\end{equation}
and
\begin{equation}
\frac{m^*(\omega)}{m_b} = 1+ \frac{\lambda_0}{1+\alpha^2\omega^2},
\end{equation}
where $\Gamma_0$ and $\lambda_0$ are the zero frequency scattering rate
and mass enhancement, respectively. These expressions obey the
Kramers-Kronig relation and have the proper FL frequency
dependence. We have performed a fit to our 20K data on (TMTSF)$_2$PF$_6$
using these forms of $\Gamma_1$ and $m^*$ and Eq.~(\ref{eq:gen-drude2}).
This result is shown by the lines with open circles on
Figs.~\ref{fig:abs-low} and \ref{fig:sig-low}, with
$\omega_p/2\pi c=1000$~cm$^{-1}$, $\Gamma_0/2\pi c=0.56$~cm$^{-1}$, 
$(2\pi c \alpha)^{-1}=1$~cm$^{-1}$ 
and $\lambda_0=1$. It is clear from the absorptivity that this is a
significantly better representation of the measured data than either of
the two simple Drude fits shown by the dashed lines.
In addition, above 0.5~cm$^{-1}$ this fit is in
excellent agreement with the conductivity obtained from the
Kramers-Kronig calculation. The differences at lower frequencies are a
result of the perfectly straight interpolation between the microwave
absorptivity points, shown by the solid line. It is clear that within
the error bars, the FL fit is an equally good interpolation.
It should also be pointed
out that the results are shown on a logarithmic frequency scale, which tends
to enhance the low frequency range. The inset of Fig.~\ref{fig:sig-low}
displays the Kramers-Kronig calculation and the fits on a linear frequency
scale, showing the obvious improvement obtained with the generalized Drude
fit. 

This analysis, however, leads to an anomalously small value for 
$1/\alpha$, as frequency dependent scattering is not expected at 
frequencies significantly less than the temperature 
($k_B T/hc \approx 14$cm$^{-1}$ at 20K). In addition,
Ruvalds and Virosztek have shown that nesting of the 
Fermi surface modifies the electron--electron scattering and leads to a 
scattering rate which is instead linear in both frequency and 
temperature. \cite{ruvalds_nested_fl} Because of the nesting of the Fermi 
surface in the (TMTSF)$_2 X$ compounds, \cite{jerome_revue_1d} we 
might expect such a form to apply in 
this case, however our attempts to fit the (TMTSF)$_2$PF$_6$ data with 
their form have proven to be unsatisfactory, giving only a small 
improvement over the simple Drude fits.
Further experiments at low frequencies could clarify these issues.

\section{Discussion}
\label{sec:disclut}

The behavior of the optical conductivity described above raises several 
interesting questions. Previous attempts to interpret some features of 
the TMTSF and TMTTF materials in terms of one-dimensional systems close 
to a Mott transition were confined to a qualitative description of the 
differences between these two families. \cite{emery_umklapp_dimerization} 
It was generally argued that only the half-filling umklapp is important, 
and that the differences between the TMTTF and the TMTSF families are due 
to the much smaller dimerization in the latter case, leading to a smaller 
$g_{1/2}$, and to the observed metallic dc behavior. The quarter-filling 
umklapp was not considered because it was beyond the reach of the 
perturbative techniques used at that time. Now, non-perturbative 
calculations of the frequency dependence of the conductivity 
\cite{giamarchi_umklapp_1d,giamarchi_mott_shortrev} enable us to 
investigate the consequences of attributing the Mott gap to either of 
these two umklapps.

If one interprets the high frequency behavior in terms of a 
one-dimensional Mott insulator, two possibilities arise depending on 
whether the half- or quarter-filled umklapp is dominant. In the former 
case, one would have $5-4K_\rho \sim 1.3$ leading to a rather weak 
repulsive system with $K_\rho \sim 0.925$. The standard perturbative 
formula, $K_\rho \simeq 1 - U/(\pi v_F)$, leads to relatively small 
interactions of strength $U/v_F \sim 0.23$. Such a small value of the 
interaction would pose several problems. It is in disagreement with 
simple estimates based on quantum-chemistry calculations 
\cite{penc_numerics} or estimations of the interactions in the spin 
sectors \cite{bourbonnais_rmn}(there is no a priori reason to assume that 
interactions would cancel in the charge sector). In addition, since the 
gap opened by the half-filled umklapp would be given by
\begin{equation} \label{legap}
\Delta_\rho \sim W \left(\frac{g_{1/2}}{W} \right)^{1/(2-2K_\rho)},
\end{equation}
the rather small half-filled umklapp constant $g_{1/2}\sim W (D/W)$, 
where $D$ is the dimerization gap, and the rather large exponent 
$1/(2-2K_\rho)\approx 6.6$ would lead to an exceedingly small value for the 
Mott gap, incompatible with the observed peak in the optical conductivity 
of Figs.~\ref{fig:pf6-as}--\ref{fig:clo4-as}.

Thus the Mott gap seen in the optical conductivity is unlikely to be
due to the half-filling umklapp. As suggested in the context of
the temperature dependence of the conductivity, 
\cite{giamarchi_mott_shortrev} a way out of this problem is to
assume that the conductivity is dominated by the quarter-filled umklapp.
In that case the exponent of the frequency dependence is $5 - 16 K_\rho
\approx 1.3$, leading to a much smaller value of $K_\rho \approx 0.23$.
Such a small value of $K_\rho$ corresponds to a relatively large
repulsion. The existence of such large repulsion is consistent with the
initial hypothesis of predominance of the quarter-filling umklapp. 
In that case $g_{1/4}/W$ can be of
order one since $g_{1/4} \sim W (U/W)^3$, whereas $g_{1/2}$, being
due to dimerization, would still be much smaller than the typical
interaction $U$ (by a factor $D/W \sim 10^{-2}$).
Such a small value of $K_\rho$ also gives a density of states exponent
$\alpha = (K_\rho+K_\rho^{-1})/4 -1/2 \approx 0.64 $, in reasonable agreement with
photoemission observations. \cite{dardel_photoemission_tmtsf,grioni_photoemission}
However the photoemission results should be taken with some degree of care, 
due to the large range of energy over which the TLL
behavior was observed. The Mott gap due to
the quarter-filled umklapp would be given by an analogous formula to
Eq.~(\ref{legap}) with $g_{1/2} \to g_{1/4}$ and $(2-2K_\rho) \to (2 - 8
K_\rho)$. The numerical value of the exponent would remain unchanged, but
because the interactions are now allowed to be larger, $g_{1/4}$ can
be closer to $W$, allowing for reasonably large values of the Mott gap.
It is of course not obvious that one can get such a small
value of $K_\rho$ for the TMTSF family from reasonable microscopic 
interactions. This is probably not a serious problem, however, 
since the range of the interaction is more important than the
strength in getting a small $K_\rho$ (a purely local repulsion 
cannot get below $K_\rho \sim 0.5$). Although a direct calculation of
$K_\rho$ from a microscopic model
is too strongly dependent on the precise details of the model to be really
quantitative, the order of magnitude of an interaction needed to get
$K_\rho\approx 0.25$ is not incompatible with what is expected
microscopically, \cite{penc_numerics} or estimated from the uniform
susceptibility. \cite{bourbonnais_rmn}

The interpretation of the optical conductivity in terms of 
quarter-filling umklapp, although quite different from the standard view, 
\cite{emery_umklapp_dimerization} thus seems more reasonable. However it 
has also its share of problems and raises interesting issues. First, if 
the quarter-filling umklapp plays an important role in the TMTSF family 
it is likely to also play a role in the parent TMTTF family. The 
dimerization gap differs by a mere factor of two between the two 
families, and it would be surprising if this simple enhancement of the 
$g_{1/2}$ constant would make it dominate over $g_{1/4}$ (which is not 
expected to change much).

The question is of importance because, until now, the different behavior of 
the dc conductivity between the two families has  been attributed to the 
change in the dimerization. \cite{emery_umklapp_dimerization} The 
reduction of dimerization with pressure was supposed to reduce $g_{1/2}$ 
and hence the Mott gap. If the dominant umklapp process is the 
quarter-filling one this explanation cannot continue to hold because 
$g_{1/4}$ is expected to be only weakly pressure dependent. However, 
experimentally one still observes a large difference in the Mott gap 
between the two families. One possible explanation for the large change 
of gap under pressure (or by changing the family) could be the change in 
the exponent of Eq.~(\ref{legap}), i.e.\ a change of $K_\rho$ and
not so much a change of the umklapp coupling constant, as is usually 
advocated. Indeed, since $K_\rho$ is close to the value where the umklapp 
term is irrelevant, even a small change in $K_\rho$ can produce a 
relatively large variation in the gap (the gap would go to zero for 
$K_\rho = 0.25$).

The main problem of such a small value of $K_\rho$ would be its
effect on the interchain hopping. Using the standard
formula for the renormalization of the transverse hopping due to
interactions \cite{bourbonnais_couplage,boies_hopping_general}
would give an effective hopping between the chains
\begin{equation} \label{effecthop}
t_\perp^{\text{eff}} \sim t_\perp
\left(\frac{t_\perp}{W}\right)^{\frac{\alpha}{1-\alpha}}
=W \left( \frac{t_\perp}{W} \right)^{\frac{1}{1-\alpha}},
\end{equation}
where $\alpha = (K_\rho+K_\rho^{-1})/4 - 1/2$ is the density of states
exponent. A value of $K_\rho=0.23$ gives
$t_\perp^{\text{eff}}\sim W (t_\perp/W)^{2.8}$, leading to a
small crossover value, on the order of 30K, between the
one-dimensional regime and a regime where the hopping between the chains
is relevant. Such a low crossover scale was indeed the one advocated
based on both NMR \cite{bourbonnais_rmn} and magnetoresistance
experiments. \cite{behnia_transport_magnetic} However,
such a small value of the effective interchain hopping
is incompatible with several observations. 

First the temperature dependence of the dc conductivity shows a crossover 
between a roughly linear $T$ behavior, that can be interpreted in terms 
of a one-dimensional regime, to a $T^2$ regime at a scale of 150--200~K. 
\cite{jerome_organic_review} It is very natural to interpret 
such a change of behavior in terms of a crossover to a more 
two-dimensional regime, where the interchain hopping is relevant, leading 
to $t_\perp^{\text{eff}}\sim 150-200$~K. Such a value is in good agreement 
with measurements of the transverse conductivity that provide a very 
sensitive way of probing this dimensional crossover, 
\cite{mihaly_cond_transverse_tmtsf,jerome_cond_transverse} and with 
direct low temperature measurements of $t_\perp^{\text{eff}}$. 
\cite{chaikin_oscillations_tperp} Second, although it is quite difficult 
to extract the crossover scale from the optical or dc conductivity along 
the chains alone, the behavior of the optical conductivity below the Mott 
peak does not follow the prediction ($\sigma(\omega)\sim \omega^3$) 
of a purely one-dimensional model very well.
\cite{giamarchi_umklapp_1d,giamarchi_mott_shortrev} 
This again suggests a crossover towards 
two-dimensional behavior at an energy scale on the order of 
$E_{\text{cr}}\sim 100-200$~K. Third, on a more theoretical level, if the 
Mott gap was much higher than the effective interchain hopping, the 
system would remain one-dimensional 
\cite{emery_umklapp_dimerization,behnia_transport_magnetic} and thus 
insulating. \cite{giamarchi_mott_shortrev} A small value of 
$t_\perp^{\text{eff}}$ is thus again incompatible with the observed 
metallic behavior at low frequency. Paradoxically, the observed crossover 
scale $E_{\text{cr}}\sim 200$~K would be compatible with 
Eq.~(\ref{effecthop}), if the value of $K_\rho$ corresponding to the 
half-filling umklapp was used, namely very weak interactions, but as we 
saw above such a value of $K_\rho$ seems incompatible with other 
observations.

Below an energy scale on the order of 200~K, the system would therefore 
be in a two-dimensional regime. The analysis of 
Sec.~\ref{sec:zero-energy}, suggests the possibility that at low frequencies and 
temperatures the behavior in this regime is that of a simple FL. 
Such behavior is reasonable, and in agreement with the 
results of NMR measurements at low temperatures. \cite{bourbonnais_rmn} 
However, the precise nature of 
the phase below the 1D to 2D crossover temperature still remains to 
be clarified, in particular at intermediate temperatures and
frequencies. Proposals have been made about the existence of a two-dimensional
Luttinger liquid with power law correlation functions. \cite{anderson_2DLL}  
Whether the experimentally observed phase is of such nature remains to
be checked. Some quantities, however, show a simple exponent (e.g.\ $\sigma(T)\sim
T^{-2}$), which is quite incompatible with the suspected power-law dependence 
of the hypothetical two-dimensional Luttinger liquid.

\section{Conclusions}
\label{sec:conclusions}

In this paper we have presented the results of our measurements of 
the on-chain, metallic state electrodynamics of three Bechgaard salts, 
(TMTSF)$_2 X$. In all three cases, we find dramatic deviations from the 
simple Drude response, with the frequency dependent conductivity instead 
consisting of two distinct features: a narrow mode at low energy
containing a very small part of the spectral weight ($\sim 1\%$)
and a high energy mode centered around 200~cm$^{-1}$.
We have argued that these are the characteristics of a highly anisotropic
interacting electron system, with either a half- or quarter-filled band.
This leads to a Mott gap and, at frequencies above the effective
interchain transfer integral, to a Luttinger liquid state --- consequently we
call this a Mott-Tomonaga-Luttinger liquid.

The finite energy feature can be successfully described as the absorption 
above the Mott gap in such a one-dimensional Luttinger liquid. In 
particular, above the gap the optical conductivity behaves as a power law 
of the frequency $\sigma(\omega) \sim (1/\omega)^\nu$, in a way 
characteristic of a TLL. The exponent $\nu$ is determined 
experimentally to be $\nu \approx 1.3$. The low energy feature in the 
conductivity can be successfully described by using a frequency dependent 
relaxation rate and effective mass with quadratic frequency dependence, 
suggestive of a Fermi liquid. However, further studies are needed at these
low energies in order to rule out other mechanisms.

Comparison of the data with the TLL theory, along with 
the value of the exponent $\nu$, suggest that the dominant mechanism 
responsible for the opening of the Mott gap is the quarter-filling of the 
band. This conclusion leads to a Luttinger liquid parameter $K_\rho \approx 
0.23$, corresponding to very strong repulsion. The fact that the 1D theory 
is unable to account quantitatively for the data below the high energy 
peak seems to suggest that the crossover to a two-dimensional regime 
occurs at relatively high energies. 

Although the conductivity at low frequencies is in agreement with that 
arising from electron--electron scattering, and at high frequencies with 
predictions based on the TLL model, a host of unresolved 
questions remain. These include the precise nature of this crossover, and 
the magnitude of the effective transfer integral $t_\perp^{\rm eff}$ 
where it occurs. It also remains to be seen whether our experiments are 
compatible with studies which directly probe the Fermi surface of these 
compounds. Experiments along directions perpendicular to the chains, 
together with optical studies on the more anisotropic TMTTF compounds, 
should help to clarify these unresolved questions, and are currently 
underway.

\acknowledgments

We would like to thank S. Donovan for all of his work on this topic, B. 
Alavi for preparation of the large single crystals used in this work, B. 
Gorshunov and Y. Goncharov for their assistance with the submillimeter 
measurements, and F. Mila, P. Chaikin, D. J\'erome, M. Gabay, and C. 
Berthier for helpful discussions. L.D. and V.V. would like to thank the 
Swiss National Foundation for Scientific Research for important 
financial support. The work at UCLA was supported by NSF Grant No. 
DMR-9503009. 


\end{multicols}
\end{document}